# Coulomb-bound four- and five-particle valleytronic states in an atomically-thin semiconductor

*Shao-Yu Chen,[1] Thomas Goldstein,[1] Takashi Taniguchi,[2] Kenji Watanabe[2] and Jun Yan[1,*]*

[1]Department of Physics, University of Massachusetts, Amherst, Massachusetts 01003, USA

[2]National Institute of Materials Science, 1-1 Namiki, Tsukuba, Ibaraki 305-0044, Japan

[*]Corresponding Author: Jun Yan.    Tel:   (413)545-0853    Fax:   (413)545-1691

E-mail: yan@physics.umass.edu

**Abstract:**

As hosts for tightly-bound electron-hole pairs carrying quantized angular momentum, atomically-thin semiconductors of transition metal dichalcogenides provide an appealing platform for optically addressing the valley degree of freedom. In particular, the valleytronic properties of neutral and charged excitons in these systems have been widely investigated. Meanwhile, correlated quantum states involving more particles are so far elusive and controversial. Here we present experimental evidence for valleytronic four-particle biexcitons and five-particle exciton-trions in high-quality monolayer tungsten diselenide samples. Through charge doping, thermal activation, and magnetic-field tuning measurements, we determined that the biexciton and exciton-trion optical emissions are bound with respect to the bright exciton and the trion respectively. Further, both the biexciton and the exciton-trion are intervalley bound states involving dark excitons, giving rise to emissions with large, negative valley polarizations in contrast to that of the well-known two-particle excitons. Our studies provide new opportunities for building valleytronic quantum devices harnessing high-order excitations.

Coulomb interaction mediated multi-particle bound-states are highly-correlated entities that offer appealing potentials for realizing Bose-Einstein condensation, matter-antimatter gamma-ray laser, and quantum communication and teleportation[1–3]. Many-body correlation is a fascinating topic that has attracted decades of experimental and theoretical investigation. In atomic physics, the positronium molecule, consisting of two electrons and two positrons bound together, took more than half a century to confirm after the discovery of the positronium atom[4]. In condensed matter systems where the positrons are replaced by holes (missing electronic states in the valence band), *ab initio* simulations are so far lacking for addressing complexes with three or more charged particles; and experimentally light emissions due to exciton molecules, or biexcitons, are only revealed in a few systems[5,6].

In this work, we report the experimental observation of optical features due to biexcitons confined to a two-dimensional semiconductor, the monolayer tungsten diselenide ($WSe_2$). Interestingly we discovered that the biexciton consists of a spin-zero bright exciton in one valley and a spin-one 'dark' exciton in the other. Such valley-spin configuration of the exciton molecule is unusual and has not been observed in other systems. Our samples also emit a lower-energy nonlinear feature, which we attribute to a five-particle bound state consisting of a bright trion and a dark exciton residing in two different valleys. As a result of their unique spin and valley configurations, both the biexciton and the exciton-trion produce emissions with valley polarization that has an opposite sign compared to the well-known bright exciton in a magnetic field[7,8], and the degree of the valley polarization is much larger. Our findings shed new light on many-body physics of transition metal dichalcogenides, and pave way for developing new valleytronic devices and spin-valley entangled photon sources.

**Results**

**Luminescent emission due to many-body correlated states in 1L-$WSe_2$**

Monolayer (1L) hexagonal $WSe_2$ is an atomically thin semiconductor with its low energy electronic states located near the K and K' corner points - the two distinct valleys - of the Brillouin zone, suitable for developing valleytronic devices[9]. The spin of the electrons and holes in the non-centrosymmetric system is further coupled to the valley degree of freedom, and the spin degeneracy at each individual valley is lifted by spin-orbit interaction (Fig.1a). This elegant band structure results in versatile spin-valley configurations for low-energy multi-particle bound-states in the crystal: 8 for the two-particle exciton species and 12 for the three-particle negative trion species (see Fig.1b and Supplementary Fig.S1). The excitons and trions can further couple to make four- and five-particle bound states, whose underlying physics is so far not well-understood, and is the focus of investigation in this work.

The variety of spin, valley and number of particles that can form energetically well-resolved bound states results in rich light-matter interaction effects and spectroscopic features in 1L-WSe$_2$, offering practically useful tools for accessing the quantized valley pseudospin[10]. Figure 1c shows a typical luminescence emission spectrum from our high-quality hexagonal boron nitride (hBN) sandwiched 1L-WSe$_2$ sample (optical micrograph in the inset; see Supplementary for sample preparation). We observe six intrinsic emission features from the sample in the energy range of 1.65-1.72 eV. The peak with the highest energy is the two-particle bright exciton X, consisting of an electron and a hole residing in the same valley with opposite spins (top middle subpanel of Fig.1b; note that the hole spin is opposite to that of the missing electron in the valence band). This spin-zero bound state carries opposite angular momentum of $\pm\hbar$ in the K and K' valleys, giving rise to valley-selective coupling with circularly-polarized optical excitation[9], a fascinating property that has been harnessed to demonstrate valley polarization as well as valley coherence in 1L-WSe$_2$[11]. The bright exciton can further bind an electron in the same or the opposite valley to form trions (Fig.1b, bottom subpanel). Our as-made sample is slightly electron doped, and indeed, we observe $T_1$ and $T_2$ emissions at 29 and 36meV below X, attributable to the negatively charged trions[12].

While being the most prominent and well-known optical feature in transition metal dichalcogenides[13–15], it is important to note that energetically in 1L-WSe$_2$, the bright exciton X is not the two-particle ground state of the system due to the particular spin ordering of the conduction band with respect to that of the valence band (Fig.1a)[16]. Instead, the dark exciton D illustrated in Fig.1b upper-right subpanel, in which the electron and hole have the same spin and reside in the same valley, is energetically more favorable[17]. In the out-of-plane direction, the dark exciton is optically silent (hence the name). However, with finite in-plane magnetic field[18] or momentum[19] D becomes visible. While our experimental setup is in back-scattering geometry (see Supplementary), the finite collection solid angle (numerical aperture NA=0.35) and the high quality of our sample enabled us to observe this optical feature in Fig.1c, located about 40meV below X, as a result of the conduction band splitting and the distinct many-body interactions[18].

The remainder two emission features in Fig.1c, denoted as XD at 18meV and TD at 49meV below X, are assigned as the biexciton four-particle state and the exciton-trion five-particle state respectively. Figure 1d compares the intensity of X, XD and TD as a function of the incident laser power in log-log scale. In contrast to X whose intensity is nearly proportional to the incident power, both XD and TD intensities rise more steeply (black dashed and dotted lines in Fig.1c are drawn for $P \propto I$ and $P \propto I^2$ respectively). This is similar to the case of biexcitons in GaAs quantum wells and carbon nanotubes[5,6], providing a first evidence that they arise from higher order complexes in the system.

**Electrostatic tuning of the many-body states**

To gain further insights into the nature of XD and TD, we fabricated a field effect transistor (FET) device using a graphene back gate, and investigated the gate voltage dependence of its luminescence. Figure 2a displays gate voltage and emission energy mapping of the luminescence intensity at 100μW excitation power over a wide tuning range from -2 to 1V; data from -0.8 to 0V at several different excitation powers are shown in Fig.2b. It is evident from these mappings that all the emission features are highly sensitive to charge doping.

Similar to the sample in Fig.1, our FET device also has minor electron doping at $V_g = 0$V, and we observe strong $T_1$, $T_2$ and TD emissions in the absence of any gate induced charge carriers. As we remove electrons from the crystal by applying a negative gate voltage, $T_1$, $T_2$ and TD rapidly decrease in intensity while X becomes stronger as the sample becomes more charge neutral. Concomitantly, the dark exciton D and the biexciton XD also become prominent at low and high laser powers respectively for -0.6 V < $V_g$ < -0.2 V (Fig.2b, bottom and top panel). At even more negative gate voltage, D and XD disappear, and X becomes significantly broadened, accompanied by the appearance of a new emission peak at about 1.71eV attributable to the positive trion excitation, indicating that the sample is doped by holes in this gate voltage range[20]. These observations suggest that XD is a charge neutral entity while TD is associated with electron doping.

To be more quantitative, we have extracted the intensity of various emission features as a function of gate voltage in Fig.2c. XD is found to appear only when the X linewidth is narrow and its intensity scales with that of D. Combined with the fact that XD is a charge neutral nonlinear optical feature, we attribute it to be a biexciton consisting of a bright and a dark exciton. Noting that the dark exciton is the lowest energy state in the system (the intervalley version of D has the same kinetic energy, but the exchange interaction raises its energy by ~10meV above D)[18], it is thus quite reasonable to conjecture that multi-particle bound states prefer to involve D excitons at low temperatures. We rule out the possibility of the XD emission to be two D excitons bound together, since the emission energy is higher than the D exciton which would otherwise suggest a negative binding – a state that is energetically unfavorable. The assignment of XD as a charge-neutral biexciton is further supported by theoretical calculations. Several independent simulations have consistently found that the biexciton binding energy in $WSe_2$ is about 18-20meV[21–23], which agrees well with our observed XD to be ~18meV below X.

The physical nature of the TD emission peak is currently controversial. A previous study of $1L\text{-}WSe_2$ has observed a similar nonlinear emission feature in samples without hBN sandwiching, and has attributed it to the biexciton[24]. This assignment however, is being debated due to inconsistency with theoretical calculations, in particular the anomalously large binding energy[21–23]. From our data in Fig.2c, the intensity of TD follows the rising and lowering of the intensities of $T_1$ and $T_2$, suggesting that the underlying excitation is linked to the trions. In light of the fact that D is the two-particle ground state as discussed above, we conjecture that TD results from the binding of a trion with a dark exciton. Theoretically the binding energy of the exciton-trion five-particle state in $1L\text{-}WSe_2$ has been calculated to be ~12-15meV[23]. We note that if we count the TD binding energy from the X emission, the value of 49meV is three to four times too large compared to the theory. However, we believe this is not a legitimate counting since the TD complex does not involve X directly. Instead, the TD binding energy should be given by $\Delta_{TD} = E_T + E_D - E_{TD}$. We assume that the emission we observe is due to the dissociation of TD into a dark exciton and a trion, which radiatively recombines subsequently. Thus, the emission energy is given by $\hbar\omega_{TD} = E_{TD} - E_D = E_T - \Delta_{TD} = \hbar\omega_T - \Delta_{TD}$, i.e., the TD binding energy needs to be counted from the trion emission energy. Indeed, the energy separation between TD and $T_2$ is 13meV, in excellent agreement with theoretical calculations.

**Thermal activation of the biexciton and exciton-trion bound states**

The binding energies of XD and TD can be further assessed from thermal activation measurements. In Fig.3 we plot the temperature dependence of 1L-WSe$_2$ luminescence. The XD and TD peaks are found to be highly sensitive to sample heating and they disappear in the temperature range of 100 to 130K. In contrast, the neutral exciton and the negative trion emissions survive to much higher temperatures. The comparatively more robust trion emission suggests that the binding energies of both XD and TD are smaller than that of the trions, further challenging the speculation that the TD peak is bound with respect to X.

Quantitatively the temperature dependence of XD and TD intensities are impacted by both the formation and the disassociation dynamics of these highly correlated complexes. The D exciton is the lowest energy state in the system; for temperatures below 130K, we can assume that there are plenty of dark excitons in the crystal. This is reflected in the dramatic dropping of X intensity at low temperatures[17,25], as well as our observation of relatively strong dark exciton emission in a backscattering optical setup with relatively small NA. The formation process can thus be assumed to be determined by the population of the minority species, namely XD by X and TD by T in the system, which we approximate by the luminescence emission intensity of the neutral and charged excitons. By normalizing the intensity of XD and TD to the intensity of X and T respectively, we quantitatively characterize the thermal dissociation of XD and TD as a function of temperature in Fig.3c. This thermally activated disassociation can be captured by using the thermal activation equation considering only one binding energy:

$$I = \frac{I_0}{1 + Ae^{-\frac{E_b}{k_B T}}} \quad (1),$$

where $I_0$ is the intensity at 0 K, $E_b$ is the binding energy, $k_B$ is Boltzmann constant, and $A$ is a fitting constant. Using Eq.(1) to fit our experimental data, we find that the binding energy of XD and TD to be 18-23meV and 13-20meV respectively. These values are in reasonable agreement with the theoretical calculations[21–23] as well as the binding energy counting alluded above.

**Valleytronic properties of the biexciton and the exciton-trion**

The biexciton and the exciton-trion complexes that we observed possess remarkable valleytronic properties. In the presence of time reversal symmetry, the electronic states at the K and K' valleys are degenerate. This degeneracy is lifted in an out-of-plane magnetic field since electronic states in the two valleys have opposite magnetic dipole moment[7,8]. The resulting Zeeman splitting then leads to valley polarized charge distribution, similar to the imbalanced spin occupation in a magnetic material, which can be monitored by the optical response of the system to photons of opposite circular polarization by virtue of the valley-helicity optical selection rules[9].

To understand the impact of valley degeneracy breaking on the four- and five-particle states, we use linearly-polarized laser light at 2.33eV to excite our sample placed in an out-of-plane magnetic field, and collect the resultant luminescent emission in a circular-polarization resolved optical spectroscopy setup[26] (see Supplementary). Figure 4a shows the intensity map of $\sigma^-$ luminescence for magnetic fields ranging from -8 to 8 Tesla. Both XD and TD emissions obey well the valley-helicity selection rule, namely, only the K' valley electron-hole recombination is allowed to occur in the $\sigma^-$ channel, similar to X, $T_1$ and $T_2$. In contrast, this valley-helicity locking is broken for the D exciton, and both K and K' dark exciton recombination shows up in the $\sigma^-$ luminescence emission, giving rise to the cross pattern in Fig.4a not observed for the other five bound states. This observation reiterates the fact that the valley-helicity locking is for angular momentum that is perpendicular to the atomic layer[9]. In this direction, the dark exciton is a spin-one entity that cannot couple to the optical fields[19]. Instead, the observed D emission arises from radiation with momentum that is not perfectly perpendicular to the atomic layer; the projection of exciton spin and angular momentum to the light propagation direction allows for coupling of D in each valley to both $\sigma^+$ and $\sigma^-$ radiation.

The slope of the XD and TD emission energy versus the magnetic field (i.e. the Zeeman shift) can be characterized by the $g$-factor: $E_Z = g\mu_B B$ where $\mu_B = 0.058$ meV/T is the Bohr magneton. From Fig.4a, XD and TD have the same slope as X, $T_1$ and $T_2$, and $g$ = 2.17. For D however, the slope is larger, and its $g$ is 4.58. This larger value of $g$ is mostly due to the spin of the dark exciton. X is a spin-zero entity and its magnetic dipole moment is purely orbital; i.e., the composing electron and hole spin contributions cancel each other [7,8]. D on the other hand, is a spin-one bound state, and the spin contributes an additional $2\mu_B$ to its dipole moment, making its Zeeman splitting almost twice that of X, as shown in Fig.4b.

The valley-helicity locking and the Zeeman $g$ factors of XD and TD emission exclude the physical picture where XD and TD emissions arise from radiative recombination of the disassociated dark exciton with finite in-plane momentum, and provide further evidence that the radiative emission of these four- and five-particle bound states are linked to either bright excitons or trions, supporting our interpretation of their formation and disassociation process as well as their binding energy interpretation discussed above.

Another important information regarding the valleytronic properties of XD and TD is encoded in the intensity of the Zeeman-split peaks. The off-resonance laser excitation with linear polarization we use populates both valleys of 1L-WSe$_2$ equally with electrons and holes. However due to the breaking of valley degeneracy, the formation probability of multi-particle bound states in the two valleys are non-equal and occupation of lower energy states is preferred. The emission intensities from different channels thus reflect the degree of valley polarization of the corresponding underlying excitonic species.

Figure 5 plots the spectral intensities of X, D, XD and TD in $\sigma^+$ and $\sigma^-$ channels at 8T. For X we observe that the $\sigma^+$ emission at K valley is more intense than $\sigma^-$ at K': this is understandable since the K valley bright exciton has lower energy at positive magnetic fields. For XD and TD in Fig.5c&d, we also observe that the $\sigma^+$ emissions have lower energy than the $\sigma^-$, confirming again the origin of the radiatively recombined electron and hole. What is unusual is their intensities: the lower energy $\sigma^+$ emissions for XD and TD are significantly weaker than the higher energy $\sigma^-$ emissions. This somewhat counter-intuitive observation is a manifestation that XD and TD are intervalley complexes, as we discuss below.

Following the convention established by previous studies[13–15], we quantify the valley polarization by:

$$P_V = \frac{I^{\sigma^+} - I^{\sigma^-}}{I^{\sigma^+} + I^{\sigma^-}} \qquad (2),$$

where the $I^{\sigma^+}$ and $I^{\sigma^-}$ are the integrated emission intensity from the $\sigma^+$ and $\sigma^-$ channels respectively. Using Eq.(2), we find $P_V$ to be 0.05 for X, -0.28 for XD and -0.33 for TD. The negative $P_V$ indicates that the dark excitons involved in XD and TD reside in the opposite valley from that of X and T; see the schematic illustrations in Fig.5e. In the presence of a magnetic field, there exist more lower Zeeman D excitons, which due to the intervalley nature of XD and TD, necessarily bind to the higher Zeeman X and T. As we discussed above, the radiation process of XD and TD involves the disassociation of the D exciton, and the X and T left behind then radiatively recombine. Hence the higher energy XD and TD emissions are more intense.

We note that using Eq.(2) for the dark exciton in Fig.5b, one would obtain a zero $P_V$. This reflects that in the absence of valley-helicity locking, the holy grail of transition metal dichalcogenide valleytronics[9], the valley degree of freedom becomes hard to access optically. Nevertheless, the different intensities of the lower energy and higher energy Zeeman peaks $I^L$ and $I^H$ still reflect the population difference of the Zeeman split dark excitons, and as expected $I^L$ is larger than $I^H$, similar to the bright exciton X. If we define the dark exciton valley polarization as $P_V' = \frac{I^L - I^H}{I^L + I^H}$, we find $P_V'$ to be 0.5. This value is much larger than the 0.05 $P_V$ for X, due to the larger Zeeman splitting of D, as well as the absence of Maialle-Silva-Sham intervalley exchange interaction[27] that has been shown to cause valley depolarization of X [25]. It is also interesting to note that $P_V'$ is larger, but reasonably close to the absolute value of $P_V$ for XD and TD. This provides yet another evidence that D is involved in XD and TD that we observe, and its valley distribution plays a dominant role in the large valley polarization of the four- and five-particle states, as compared to the bright excitons.

In conclusion, we observed six intrinsic low-energy emission features arising from bound quantum states in 1L-WSe$_2$. The presence of strong Coulomb interaction and the high quality of our sample enabled observation of the four-particle XD and five-particle TD bound states under a non-resonant continuous wave excitation. We assign XD as the intervalley biexciton composed of a spin-1 dark exciton and a spin-0 bright exciton, and TD as the intervalley exciton-trion consisting a spin-1 dark exciton and a negatively charged trion. Luminescence measurements at finite magnetic fields reveal the unusual negative valley polarization for the XD and TD emission, highlighting the role of dark excitons in forming the multi-particle bound states and their intervalley nature. Our results reveal rich many-body correlated excitonic physics and pave way to novel applications such as those involving valley encoded quantum information.

**Data availability**

The data that support the findings of this study are available from the corresponding author on reasonable request.

**Acknowledgements**

This work is supported mainly by the University of Massachusetts Amherst, and in part by NSF ECCS-1509599. T.T. and K.W. acknowledge support from the Elemental Strategy Initiative conducted by the MEXT, Japan and JSPS KAKENHI Grant Number JP15K21722. We thank Dr. Jiayue Tong for helping with e-beam lithography to make the graphene back-gated FET device.

**Author contributions**

S.-Y.C. and J.Y. conceived the optical and magneto-optical experiments. T.G. performed the CVT growth of bulk WSe$_2$ crystal. T.T and K.W. grew the single crystal of bulk hBN. S.-Y.C. fabricated the BN/1L-WSe$_2$/BN heterostructures. S.-Y.C. performed the gate-dependent, temperature-dependent and magnetic field-dependent luminescence measurements. S.-Y.C. and J.Y. co-wrote the paper. All authors discussed the results, edited, commented and agreed on the manuscript.

**Competing financial interests**

The authors declare no competing financial interests.

**References**

1. Blatt, J. M., Böer, K. W. & Brandt, W. Bose-Einstein Condensation of Excitons. *Phys. Rev.* **126,** 1691–1692 (1962).
2. Mills, A. P., Cassidy, D. B. & Greaves, R. G. Prospects for Making a Bose-Einstein-Condensed Positronium Annihilation Gamma Ray Laser. *Mater. Sci.*

*Forum* **445–446,** 424–429 (2004).

3. Walls, D. F. & Milburn, G. J. (Gerard J. . *Quantum optics*. (Springer, 2008).
4. Cassidy, D. B. & Mills, A. P. The production of molecular positronium. *Nature* **449,** 195–197 (2007).
5. Colombier, L. *et al.* Detection of a Biexciton in Semiconducting Carbon Nanotubes Using Nonlinear Optical Spectroscopy. *Phys. Rev. Lett.* **109,** 197402 (2012).
6. Kim, J. C., Wake, D. R. & Wolfe, J. P. Thermodynamics of biexcitons in a GaAs quantum well. *Phys. Rev. B* **50,** 15099–15107 (1994).
7. Srivastava, A. *et al.* Valley Zeeman effect in elementary optical excitations of monolayer WSe2. *Nat. Phys.* **11,** 141–147 (2015).
8. Aivazian, G. *et al.* Magnetic control of valley pseudospin in monolayer WSe2. *Nat. Phys.* **11,** 148–152 (2015).
9. Xiao, D., Liu, G., Feng, W., Xu, X. & Yao, W. Coupled spin and valley physics in monolayers of MoS 2 and other group-VI dichalcogenides. *Phys. Rev. Lett.* **108,** 196802 (2012).
10. Xu, X., Yao, W., Xiao, D. & Heinz, T. F. Spin and pseudospins in layered transition metal dichalcogenides. *Nat. Phys.* **10,** 343–350 (2014).
11. Jones, A. M. *et al.* Optical generation of excitonic valley coherence in monolayer WSe2. *Nat. Nanotechnol.* **8,** 634–8 (2013).
12. Jones, A. M. *et al.* Excitonic luminescence upconversion in a two-dimensional semiconductor. *Nat. Phys.* **12,** 323–327 (2015).
13. Zeng, H., Dai, J., Yao, W., Xiao, D. & Cui, X. Valley polarization in MoS2 monolayers by optical pumping. *Nature Nanotechnology* **7,** 490–493 (2012).
14. Mak, K. F., He, K., Shan, J. & Heinz, T. F. Control of valley polarization in monolayer MoS2 by optical helicity. *Nature Nanotechnology* **7,** 494–498 (2012).
15. Cao, T. *et al.* Valley-selective circular dichroism of monolayer molybdenum disulphide. *Nature Communications* **3,** 887 (2012).
16. Liu, G.-B., Shan, W.-Y., Yao, Y., Yao, W. & Xiao, D. Three-band tight-binding model for monolayers of group-VIB transition metal dichalcogenides. *Phys. Rev. B* **88,** 85433 (2013).

17. Zhang, X.-X., You, Y., Zhao, S. Y. F. & Heinz, T. F. Experimental Evidence for Dark Excitons in Monolayer WSe2. *Phys. Rev. Lett.* **115,** 257403 (2015).

18. Zhang, X.-X. *et al.* Magnetic brightening and control of dark excitons in monolayer WSe2. *Nat. Nanotechnol.* **12,** 883–888 (2017).

19. Wang, G. *et al.* In-Plane Propagation of Light in Transition Metal Dichalcogenide Monolayers: Optical Selection Rules. *Phys. Rev. Lett.* **119,** 47401 (2017).

20. Hao, K. *et al.* Direct measurement of exciton valley coherence in monolayer WSe2. *Nat. Phys.* **12,** 677–682 (2016).

21. Zhang, D. K., Kidd, D. W. & Varga, K. Excited Biexcitons in Transition Metal Dichalcogenides. *Nano Lett.* **15,** 7002–7005 (2015).

22. Mayers, M. Z., Berkelbach, T. C., Hybertsen, M. S. & Reichman, D. R. Binding energies and spatial structures of small carrier complexes in monolayer transition-metal dichalcogenides via diffusion Monte Carlo. *Phys. Rev. B* **92,** 161404 (2015).

23. Kylänpää, I. & Komsa, H.-P. Binding energies of exciton complexes in transition metal dichalcogenide monolayers and effect of dielectric environment. *Phys. Rev. B* **92,** 205418 (2015).

24. You, Y. *et al.* Observation of biexcitons in monolayer WSe2. *Nat. Phys.* **11,** 477–481 (2015).

25. Chen, S.-Y. *et al.* Superior Valley Polarization and Coherence of 2s Excitons in Monolayer WSe2. *Phys. Rev. Lett.* **120,** 46402 (2018).

26. Chen, S.-Y., Zheng, C., Fuhrer, M. S. & Yan, J. Helicity resolved Raman scattering of MoS2, MoSe2, WS2 and WSe2 atomic layers. *Nano Lett.* **15,** 2526–2532 (2015).

27. Maialle, M. Z., de Andrada e Silva, E. A. & Sham, L. J. Exciton spin dynamics in quantum wells. *Phys. Rev. B* **47,** 15776–15788 (1993).

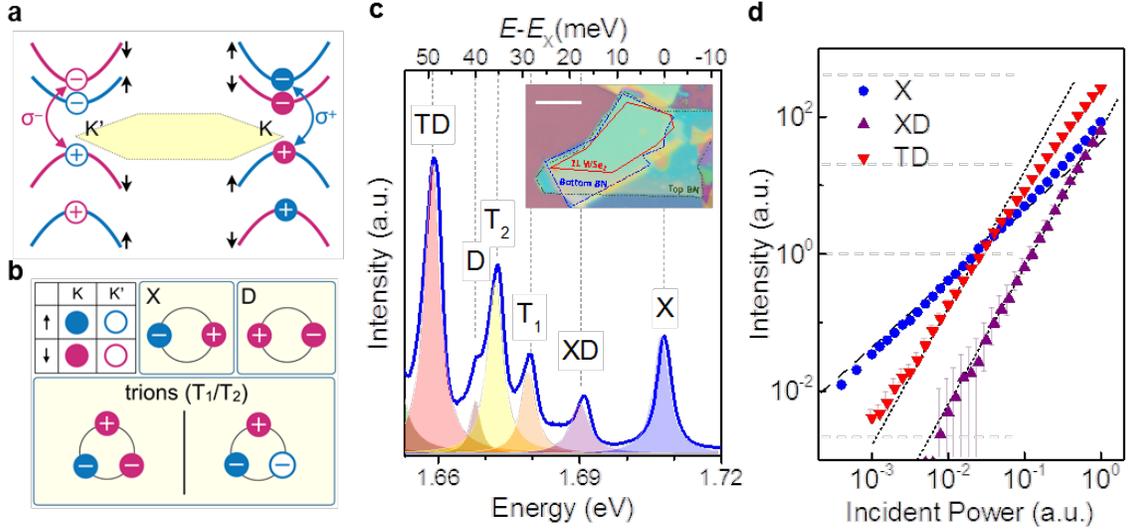

**Figure 1 | Multi-particle bound states in 1L-WSe$_2$**. **a.** Conduction and valence band configurations at the K and K' valleys. **b.** The schematics of valley-spin configurations of bound states. We use blue (red) to denote spin up (down) and closed (open) symbols to represent K (K') valley. Illustrated are the bright exciton, the dark exciton and the negatively charged trions. Only states with holes in the K valley upper valence band are shown here. **c.** A typical luminescence spectrum of high-quality 1L-WSe$_2$ excited by 2.33 eV laser at 3K. Inset: The OM image of the sandwiched BN/1L-WSe$_2$/BN heterostructure. The scale bar is 10 $\mu m$. **d.** The power dependent intensity of X, XD and TD emissions. The dashed (dot) lines in the figure are drawn for $P \propto I$ ($P \propto I^2$).

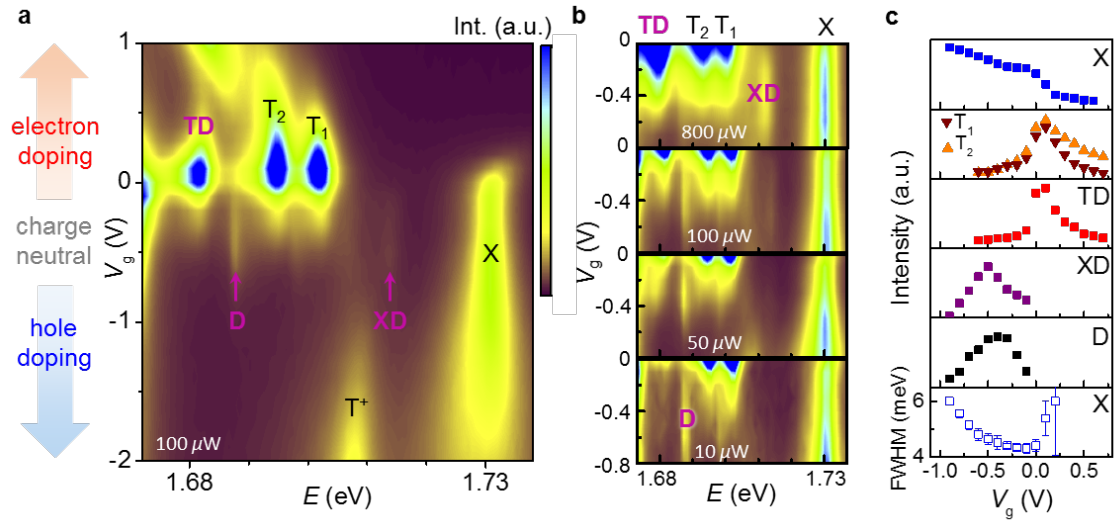

**Figure 2 | Charge doping control of the TD and XD multi-particle states. a.** Color map of PL spectra at 3K plotted as a function of gate voltage. **b.** The zoom-in color maps of PL spectra near the charge neutral region under various excitation power. **c.** Gate dependent intensity of corresponding peaks denoted in **a** and **b**. The bottom subpanel shows the gate dependent FWHM of X.

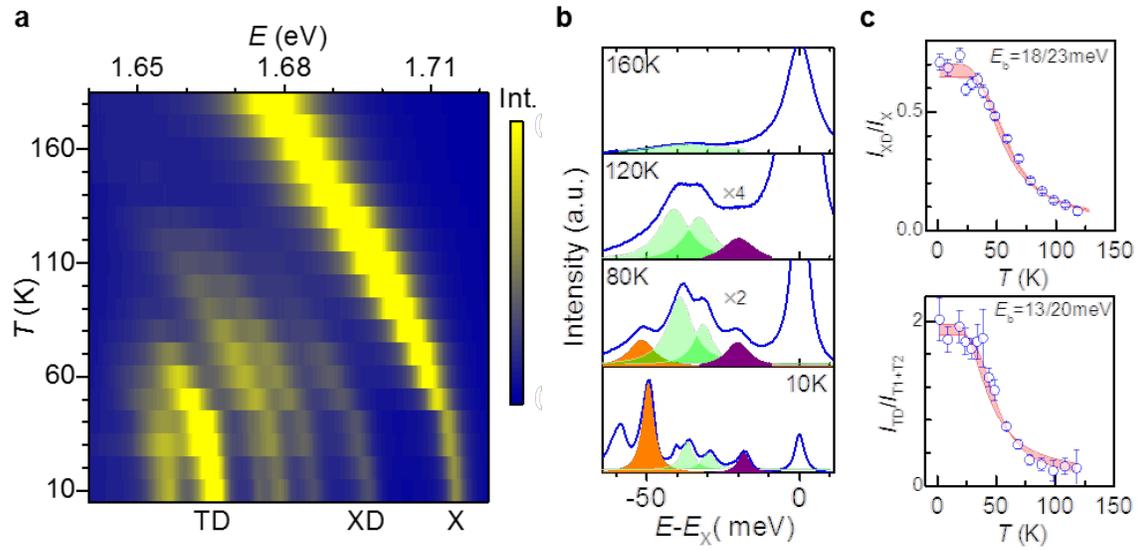

**Figure 3 | Temperature dependence of the excitonic emissions. a.** The color map of PL spectra for temperatures ranging from 3K to 180K. **b.** Selected spectra at different temperatures showing the evolution of XD (purple) and TD (orange) emissions. **c.** The normalized intensity of XD and TD plotted as a function of temperature. The decrease of intensity reveals the thermally activated dissociation as determined by the binding energy.

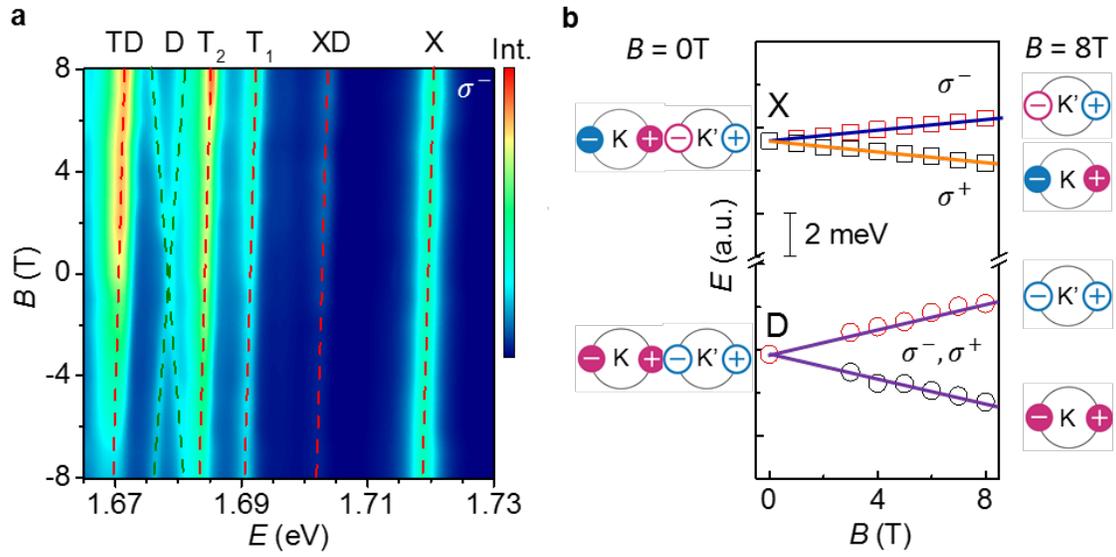

**Figure 4 | Magnetic field tuning of the XD and TD emissions. a.** The color map of $\sigma^-$ PL spectra in a perpendicular magnetic field from -8 to 8 Tesla. **b.** Magnetic field dependence of X and D Zeeman splitting. The K' valley states, degenerate with K states at zero fields, have higher energies at B = 8T. The symbols are extracted from data in panel a. Note that the X emissions are valley-helicity locked, but the D emissions are unlocked.

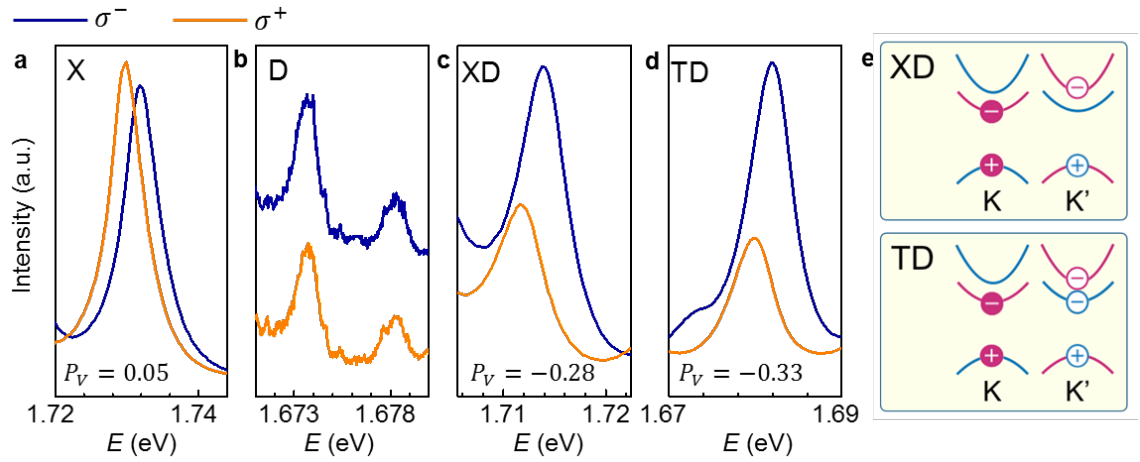

**Figure 5 | The valley polarization of excitonic bound states in a finite out-of-plane magnetic field. a-d.** The luminescence spectra of X, D, XD and TD emissions in $\sigma^+$ and $\sigma^-$ channels at 8 Tesla. Valley polarization is defined as $P_V = (I^{\sigma^+} - I^{\sigma^-})/(I^{\sigma^+} + I^{\sigma^-})$. **e.** The schematics of the spin-valley configuration of the XD and TD bound states.

# Supplementary information:

1. **Spin-valley Configurations of low-energy two- and three-particle bound states in 1L-WSe$_2$**

    As illustrated in Fig. 1a in the main text, the breaking of inversion symmetry as well as the strong spin-orbital coupling lifts the spin degeneracy for both the conduction and the valence band of 1L-WSe$_2$. The splitting in the valence band is much larger than that in the conduction band (about one order of magnitude), about a few hundred meV[1]. In this paper, we focus on low-energy excitonic excitations that do not involve the lower valence band. To simplify the discussion, in this section we only show the three-band configuration as displayed in Fig.S1a. The notations of spin and valley follow the convention used in the main text: blue and red represent spin up and down, and filled and open symbols represent K and K' valleys respectively. Figures S1b to S1d illustrate all possible configurations of 2- and 3-particle states in 1L-WSe$_2$. Restricting the hole to be located at the K valley, there are in total 4 possible configurations for the 2-particle neutral exciton states, and 6 for the 3-particle negative trion states. Swapping particles in K and K' valleys, we obtain 8 exciton and 12 trion configurations. For the positively charged trions, there is one hole in each valley, and there are four options for the spin and valley choice of the electron. The enlisted states are at least doubly degenerate due to the time reversal symmetry. By applying a finite perpendicular magnetic field, the valley degeneracy can be lifted as discussed in the main text.

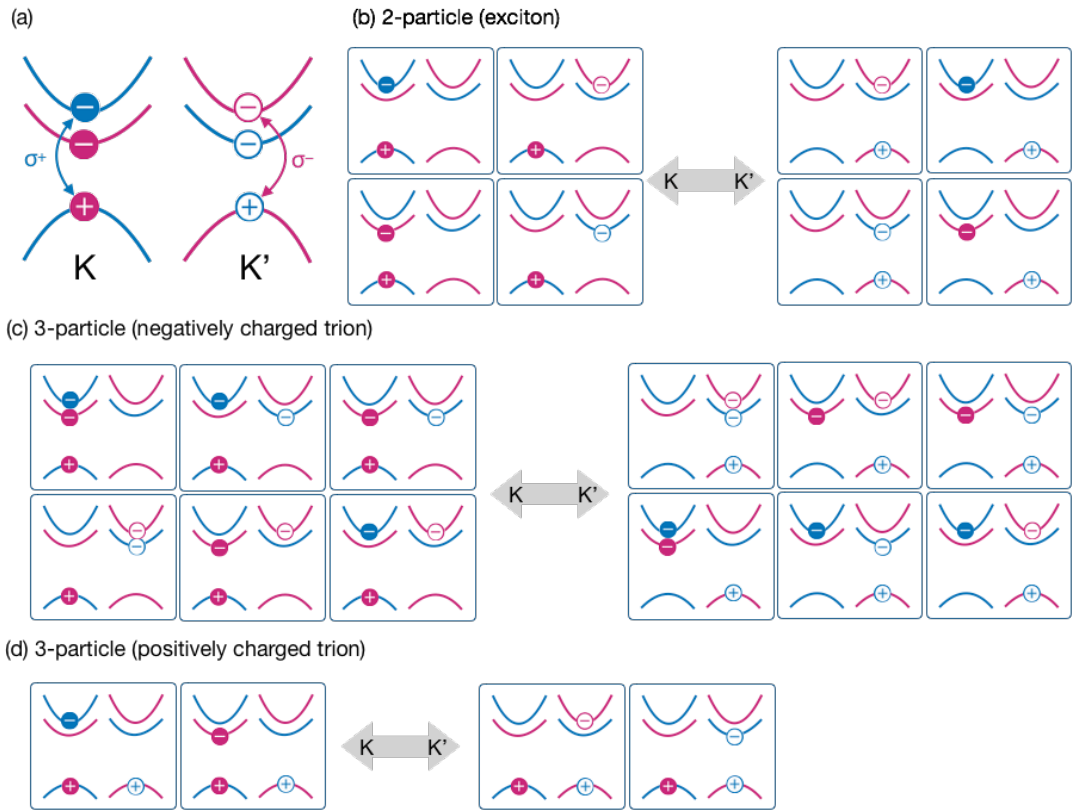

Figure S1 (a) The three-band spin-valley configurations of 1L-WSe$_2$ encoded by different colors and symbols: blue (red) for spin up (down); closed (open) for K (K') valley. There are in total (b) 8 configurations of 2-particle exciton; (c) 12 configurations of 3-particle negatively charged trions; and (d) 4 configurations of 3-particle positively charged trions.

2. **Sample Preparation**

The bulk $WSe_2$ crystals are grown by the chemical vapor transport (CVT) method. High purity W 99.99%, Se 99.999%, and $I_2$ 99.99% (Sigma Aldrich) are placed in a fused silica tubing that is 300 mm long with an internal diameter of 18 mm. W and Se are kept in a 1:2 stoichiometric ratio with a total mass of 2g. Sufficient $I_2$ is added to achieve a density of 10 mg/cm$^3$. The tube is pump-purged with argon gas (99.999%) and sealed at low pressure prior for growth. Using a three zone furnace, the reaction and growth zones are set to 1055°C and 955°C, respectively.

The atomic flakes of $WSe_2$, hexagonal boron nitride (hBN) and graphene (for making the FET sample) are first exfoliated on Si wafers with 300 nm of $SiO_2$ and inspected under optical microscope. To make the high quality 1L-$WSe_2$ heterostructures, we further apply differential interference contrast (DIC) microscopy and atomic force microscopy to select the residue-free flakes to achieve the best quality. Figure S2 shows the typical micrographs taken by DIC microscopy. Figure S2a is an exfoliated few layer hBN. The tape residues can be clearly spot to find the clean and flat hBN for sample encapsulation. Figure S2b show a DIC micrographs of the BN/1L-$WSe_2$/BN heterostructure. The sandwiched 1L-$WSe_2$ can be easier identified which is usually difficult for conventional optical microscope. The screened flakes are then stacked using a dry transfer technique with PPC (poly-propylene carbonate) stamp[2]. All the exfoliation, inspection and stacking processes are completed in a nitrogen purged dry glovebox to minimize sample degradation. The sandwiched sample is thermally annealed at 350 °C for 1 hour in argon environment to improve the quality. For the gated sample, in the stacking process, we further add a few-layer graphene as a back-gate electrode to tune the carrier density in the 1L-$WSe_2$, using hBN as a dielectric.

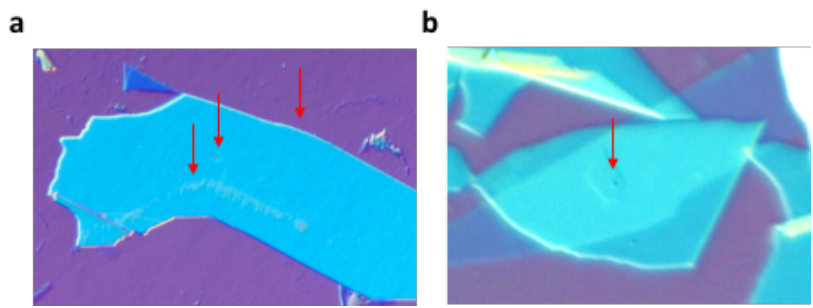

**Figure S2** (a) The DIC images of typical exfoliated few layer hBN on a $SiO_2$/Si substrate. The tape residue on top and edge of the hBN flake is indicated. (b) DIC image of BN/1L-$WSe_2$/BN heterostructure. The arrow points to the tiny bubble trapped on the 1L-$WSe_2$.

## 3. Optical Measurement Setup

The sample is transferred to a closed-loop cryostat (Montana Instruments) with a base temperature of 3K for spectroscopy measurements. The optical setup is similar to the ones used in our previous work[3]. The incident laser at 2.33 eV is focused on the sample by a 50× objective lens (NA: 0.35) with a spot size of ~2 $\mu$m. To avoid sample heating, we keep the power less than 100 $\mu$W. However, when demonstrating the nonlinearity of peak intensity as shown in Figure 1c in the main text, the heating effect is not negligible with the excitation power higher than 100 $\mu$W, which is reflected in the linewidth broadening. The luminescence signal is detected by a triple spectrometer (Horiba T64000) equipped with a liquid nitrogen cooled CCD camera.

For magneto-optical measurement, we integrate a cryostat with optical access and a 9T superconducting magnet with a room temperature bore as illustrated in Figure S3. To achieve the equal population of K and K' valleys, we excite the sample with linearly polarized light at 2.33eV. In the collection path, we first employ a quarter waveplate to transform the $\sigma^-$ and $\sigma^+$ helicity light into linearly polarized light with perpendicular polarization, followed by a half waveplate and a linear polarizer to resolve the two linear polarized signals.

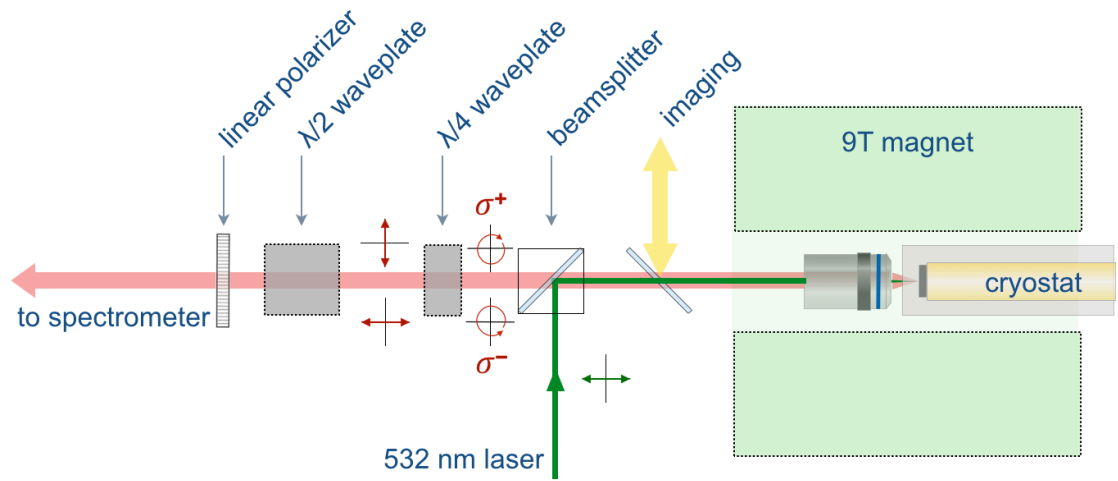

**Figure S3.** The experimental setup of valley polarization measurements.

## 4. $\sigma^+$ Helicity Luminescence in Magnetic Field

Complementary to the Figure 4a in main text, in Fig. S4a we show the color map of magnetic field dependent PL spectra in $\sigma^+$ helicity. The opposite behaviors in energy shift and intensity profiles are consistent with the discussion in the main text, reflecting the time reversal of $\sigma^-$ helicity luminescence.

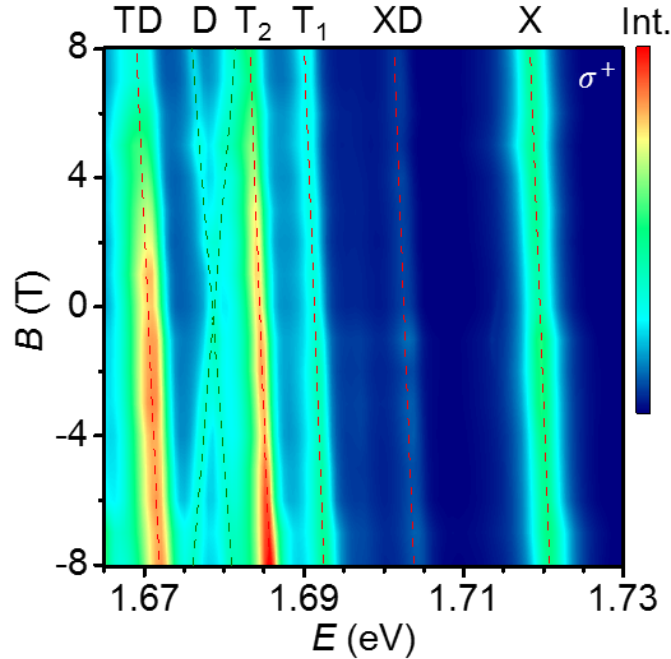

**Figure S4.** The color map of $\sigma^+$ PL spectra excited at 2.33eV in a perpendicular magnetic field from -8 to 8 Tesla.

References:
1. Liu, G.-B., Shan, W.-Y., Yao, Y., Yao, W. & Xiao, D. Three-band tight-binding model for monolayers of group-VIB transition metal dichalcogenides. *Phys. Rev. B* **88,** 85433 (2013).
2. Wang, L. *et al.* One-dimensional electrical contact to a two-dimensional material. *Science* **342,** 614–617 (2013).
3. Chen, S.-Y., Zheng, C., Fuhrer, M. S. & Yan, J. Helicity resolved Raman scattering of MoS2, MoSe2, WS2 and WSe2 atomic layers. *Nano Lett.* **15,** 2526–2532 (2015).